\documentclass[12pt]{article}
\usepackage{amstex}
\usepackage{amssymb}

\usepackage{epsfig}

\usepackage{cite}
\begin{document}
\def\Was{W\c as}
\def\Order#1{${\cal O}(#1$)}
\def\lint{\int\limits}
\def\bbeta{\bar{\beta}}
\def\tbeta{\tilde{\beta}}
\def\talpha{\tilde{\alpha}}
\def\tomega{\tilde{\omega}}
 
\begin{titlepage}

\begin{flushright}
{\bf  CERN-TH/97-162  }
\end{flushright}

\begin{center}
{\bf\LARGE
Trefoil knot and \\
\vskip 2 mm
ad-hoc classification of  elementary fields \\
\vskip 2 mm
in the Standard Model
}
\end{center}

\vskip 1 cm 
 
\begin{center}
   {\bf Z. W\c{a}s$^{\star,\dag}$ }\\
  {\em CERN, Theory Division, 1211 Geneva 23, Switzerland,}\\
   and \\
   {\em Institute of Nuclear Physics,
        Krak\'ow, ul. Kawiory 26a, Poland}\\
\end{center}
 
\vskip 4 cm

\begin{abstract}

We present an arbitrary model based on the trefoil knot 
to construct objects of the same spectrum as 
that of elementary particles. It includes `waves' and three identical 
sets of sources. Due to Lorentz invariance, `waves' group into 3 
types of 1, 3 and 8 objects and  `sources'
consists of 3 identical sets of 30+2 elements, which separate into:
$1 \times 1 \times 2 + 1 \times 2 \times 2 + 3 \times 2 \times 2 +
3 \times 1 \times 2 + 3 \times 1 \times 2 $ and another
 $1 \times 1 \times 2$ group (which does not match classification
of the Standard Model fields). On the other hand, there is no room
in this construction 
for objects directly corresponding to Higgs-like degrees of freedom.

\end{abstract}

\vskip 2 cm
 

\vskip 0.5 cm 
 
\vspace{0.3cm}
\renewcommand{\baselinestretch}{0.1}
\footnoterule
\noindent
{\footnotesize
\begin{itemize}
\item[${\dag}$]
Work supported in part by Polish Government grants KBN 2P30225206,
2P03B17210, 
\item[${\star}$]
{\tt www} home page at {\tt http://hpjmiady.ifj.edu.pl/}
%
\end{itemize}
}
\renewcommand{\baselinestretch}{1.0}

\vspace{0.1cm}
\begin{flushleft}
{\bf  CERN-TH/97-162 \\ July, 1997}
\end{flushleft}
 
\end{titlepage}
 
\section{Introduction }

The idea of using knots for descrbing elementary fields is not new.
It was proposed in 1867 by Lord Kelvin \cite{kelvin} for atoms. For about
20 years his theory was taken seriously. It eventually failed, but 
provided inspiration for the first extensive mathematical studies on 
knots, see e.g. \cite{Tait}.
Recently \cite{niemi}, it was found that knot structures form
stable solutions in a model proposed in \cite{fadeev}.
This model describes the 3+1 dynamics of a three-component vector
of unit length. Such a vector field is a typical degree of freedom
in the non-linear $\sigma$-model, a prototype relativistic quantum
field theory.

The mathematical theory of knots  \cite{knot} deals with knots constructed 
out of a ``closed loop of rope''. The trefoil knot (fig. 1) is the simplest 
example of such torus-knots.
In this letter, we will attempt to classify, in an ad-hoc manner,
knot-like stuctures obtained from the trefoil knot by placing it into different
positions of 3+1 --~dimensional space-time. We will discuss related open knots
as well, which can be obtained from a trefoil knot by the single cut 
(or by extending
one of its loops to infinity).
As no attempt will be made
to define any underlying theory, which could lead for example 
to solutions numbered 
by knots, our presentation will demonstrate a substantial degree
of ambiguity and arbitrarines.

Let us start with a glance at the lattice-like structure in 3+1 --~dimensional 
space. From every junction  there can be links directed into four direction:
$x$, $y$, $z$ and $t$.
Every link can ``vibrate'' along  three directions perpendicular to itself.
This leads to
$3 \times 4 = 12$ independent degrees of freedom. It is obvious 
that rotations belonging to the Lorentz group in a natural way provide a
transformation relating vibrational degrees of freedom of $x$, $y$ and $z$
links but {\it not $t$}. This leads to the possible different classification
of vibrational degrees of freedom for such links, namely
$3 \times 4 = 12 = 3+9$. Now,  let us realize that a simultaneous 
move of the endpoints of $x$, $y$ and  $z$ links along the $t$ direction 
is equivalent to an opposite shift of the central
junction under consideration  along the $t$ direction.
This leads to a possible separation of our degrees of freedom
accordingly to the following pattern:
$4 \times  3 = 12 = \bar 3 + \bar 8 + \bar 1$. It should be noted, that 
the separation $9=1+8$
is much less founded than the previous one. Of course any attempt
to draw a parallel between such classifications of vibrational degrees of 
freedom and the degrees of freedom of bosonic fields in the Standard Model of 
elementary particles must find healthy critic as arbitrary\footnote{ 
At least, the following objection is in place here:
Why, if the $t$-link is not connected  by Lorentz space-time transformations 
with  space-like links and  we put them  into separate groups, do we consider
vibrational modes of space links  in the $t$-direction and space-like 
directions as belonging to the same group?
A possible explanation is that vibration is on 
a smaller scale, 
where only topological properties such as continuity or closeness count.}.

Now, let us move to the trefoil knot. It is an object of 3 dimensions.
We can think of it as of a basically flat 2-dimensional object, as represented
in our figures. There are 3 and just 3 such orthogonal 
planes: $x-y$, $y-z$, $z-x$\footnote{ Alternatively, we can  think of it
as a 2-dimensional object spanned on  one space and one time direction:
$z-t$, $y-t$, $x-t$, or, what is equivalent, as a 3-dimensional object from 
any of three 3-dimensional sub-spaces: $x-y-t$, $x-z-t$ and  $z-y-t$
including the time direction.}.
As a consequence all objects of our further  classification will 
automatically appear in 3 identical copies\footnote{ Again, healthy critic
to any attempt to make a
parallel with the existence of 3 families of the Standard Model
is fully justified.}. We will continue now with a presentation of objects
from just one of these classes.

To  obtain the objects we will  cut the trefoil knot (see fig. 1)  and extend
the loose ends  into any direction of space time, irrespective of the 
original orientation of the knot plane\footnote{ We can expect that 
the orientation of the knot loop, defining family, is irrelevant, but ``true''
first, second and third ``family'' knots are  a kind 
of linear combinations or sort of superpositions of the basic states.
One could even start to speculate on mass hierarchy and/or family
mixing at this point.  
}.
The objects will group naturally,
depending on the Lorentz-separated zones the loose ends happen to point at.

As the first choice we take both loose ends pointing into the
time-direction. We end up with
$ 1 \times 2 \times 2$ possibilities corresponding to 
links pointing into the future or the past and  two possibilities  of  left- 
or right-handed knot (see fig 2). 
Such objects have loose ends that can be source of vibrations of
the  $ \bar 3 $ modes and the knot as a whole can vibrate with $ \bar 1$ mode. 
Well, this looks very much like 
colour-singlet  weak doublets  of
$(e, \nu_e)_L$ and $(\bar e, \bar \nu_e)_R$.

If we take one loose  end pointing into one of the three 
possible space directions,
another one into the time-direction of future or past, together with a choice
of left- or right-handed knot,
we have in total $ 3 \times 2 \times 2$ possibilities (see fig. 3).
Such objects, owing to their free links, can be  sources 
of vibrations of any of
the  $\bar 3+ \bar 8 + \bar 1$ modes. Well, it looks very much like 
$ (u,d)_L^{r,b,g} + (\bar u, \bar d)_R^{r,b,g}$ 3-colour weak doublets of
left-handed quarks and right-handed antiquarks.

If we take both loose ends pointing into the 
space-direction (see figs. 4 and 5) 
we end up with $ (3+3) \times 1 \times 2$ possibilities\footnote{
We can  argue that the two ends may either  follow the same 
direction, giving 3 possibilities, or follow orthogonal directions:
again $ {3 \cdot 2 \over 2}=3$ possibilities.}.
Such objects, owing to their free links,  can be sources of vibrations of
the  $ 0 + \bar 8 + \bar 1$ modes. Well, it looks very much like 
two separate classes for $ (u)_R^{r,b,g} + (\bar u)_L^{r,b,g}$ 
and $ (d)_R^{r,b,g} + (\bar d)_L^{r,b,g}$,  i.e. 3-colour weak singlets of
right-handed up/down-quark and left-handed  up/down-antiquark.

There are also two non-cut trefoil knots: left and right (see fig. 1).
As they span no external links at all, they  can be coupled to 
vibrational degree of freedom of type $ \bar 1$ only.
That is why we can note them as objects of the type $1 \times 1 \times 2 $.
These two objects we will associate, again in a somewhat arbitrary way, with
$e_R$ and $\bar e_L$.

At this point we are left with the complete list of the Standard Model degrees 
of freedom, except for the, 
as yet not discovered experimentally, Higgs field.
On the other hand, we still have $ 1 \times 1 \times 2$ possibilities 
of a knot structure when one of the loose ends points into the 
future and another 
one into the past (see fig. 6). 

Should the cut-knot of fig. 6 be considered as just a
kind of linear combinations of the basic states from  $ 1 \times 2 \times 2$ class? 
We may also guess that these cut-knots are  related
to  degrees of freedoms of `new physics', e.g. the Higgs field.
Please note, however that we have in total $3\times 2$, thus 
6 such knots, corresponding
respectively to left and right knot of three  `families'. On the other hand, 
we need only 4 degrees of freedom for the Higgs field in the Standard Model.

Let us now sketch another construction, but
basically equivalent to the previous one, where the function of 
loose ends and a knot
loop simply get interchanged\footnote{Here loose ends number 3 
possible families
and ``vibration'' of closed loop can be a source of ``waves''.}. 
Let us go back to our 3+1 --~dimensional lattice. If we assume that it 
is built 
out of infinite string-like lines, following the three $x$, $y$ or $z$ 
directions, crossing at junctions, then knots  (as in fig. 6) can appear 
as kind of defects. Now we can realize that the 
knot loop  may extend to  surround  neighbouring string-like lines 
or junctions in:
(i) time direction, (ii) time and space directions, (iii) space direction,
(iv) two distinct space directions, (v) no overlap at all. It is interesting 
to realize that in this way 
we end up with a spectrum of knots that coincides with the 
one presented above\footnote{ An exception is that our spurious state of 
the knot with one loose link pointing into the future and another one 
into the past, does not fit nicely  into this new representation.} in our letter. 
It thus again coincides with the spectrum of elementary fermions, 
which are again grouped into 3 families.

The constructions presented above are by no means unique. They all  rely
on the simplest knot of a ``closed loop of rope'', the trefoil knot, an
object from the mathematical theory of knots\footnote{ In fact 
they rely on basic, rather geometrical than topological properties of knots.
The following properties are important: (i)
the knot can be left-handed or right-handed;
(ii) it provides plane (or direction) which can be positioned 
into three distinct positions; 
(iii) and two 4-vectors which can be positioned into 3+1 orthogonal directions;
(iv) every 4-vector can `vibrate' into all perpendicular to itself directions.
} \cite{knot}.

I believe, however, that proliferating such ambiguous constructions
does not make much sense without attempting to construct models
for possible underlying dynamics, which could lead to predictions confrontable
to the data. So far this construction cannot even be disproved. 

Somehow I cannot make my mind up on
whether this construction is entertainment or
a sketchy  hypothesis. In any case, this construction is simple, counts
all observed elementary fields of the Standard Model and leaves very little 
 room for anything else. 

\vskip 8 mm
\newpage

{\bf Acknowledgements}

\vskip 6 mm

I would like to thank Prof. M. Veltman for useful discussions.
I would like  especially to thank him
for his initial discouraging me to publish this letter. This  helped me to 
mature
the ideas. I would like to thank him also  for asking me,  several months 
later, 
whether I had published it anyway. This encouraged me to publish it finally.

\vskip 2 cm

\centerline{ \large Figure captions}

\begin{enumerate}
\item
  Trefoil knot and its mirror image: two distinct objects.
\item
  Cut-trefoil knot; (i) both loose links pointing into
future/past, (ii) knot and its mirror image. In total
 $2 \cdot 2=4$  distinct objects.
\item
  Cut-trefoil knot: 
(i) first  loose link pointing into
future/past, (ii) second  loose link pointing into any of 3 space directions, 
(iii) knot and its mirror image.
In total:
 $3 \cdot 2 \cdot 2=12$  distinct objects.
\item
  Cut-trefoil knot: 
(i) first  loose link pointing into any of 3 space directions, 
(ii) second  loose link pointing into the same direction, no new possibilities.
(iii) knot and its mirror image.
      In total: $3 \cdot 1 \cdot 2=6$
  distinct objects. 
\item
 Cut-trefoil knot: 
(i) first  loose link pointing into any of 3 space directions, 
(ii) second  loose link pointing into a different  direction, no new 
possibilities, because of symmetrization factor ${1 \over 2}$, 
(iii) knot and its mirror image.
      In total: $3 \cdot 1 \cdot 2=6$
  distinct objects. 
\item
  Cut-trefoil knot and its mirror image; one loose end points into the 
future, the other into the past. We get just $2$ possibilities.
\end{enumerate}




\begin{thebibliography}{10}
\bibitem{kelvin}
 W. H. Thomson, Trans. R. Soc. Edin. {\bf 25} (1869) 217.
\bibitem{Tait} P.G. Tait, ``On Knots I, II, III'' Scentific Papers,
   Cambridge University Press, 1990. 
\bibitem{niemi}
 L. Faddeev, A. J. Niemi, ``Knots and Particles'', preprint
hep-th/9610193, 24 Oct 1996.
\bibitem{fadeev}
  L. Fadeev, Quantization of solitons, preprint IAS Print-75-QS70, 1975;
and in {\it `Einstein and several contemporary tendencies in the field theory
of elementary particles'} in Relativity, quanta and cosmology, vol. 1, 
M. Pantaleo and F.~De~Finis~(eds.).
\bibitem{knot} L. H. Kauffman, ``Knots and Physics'', World Scientific
Pub. Co., Singapore, 1991. 

\end{thebibliography}



\newpage

\centerline{\;}

\begin{figure}
\centering
\setlength{\unitlength}{0.03mm}
\begin{picture}(2400,3400)
\put(-500,-1500){\makebox(0,0)[lb]{
\epsfig{file=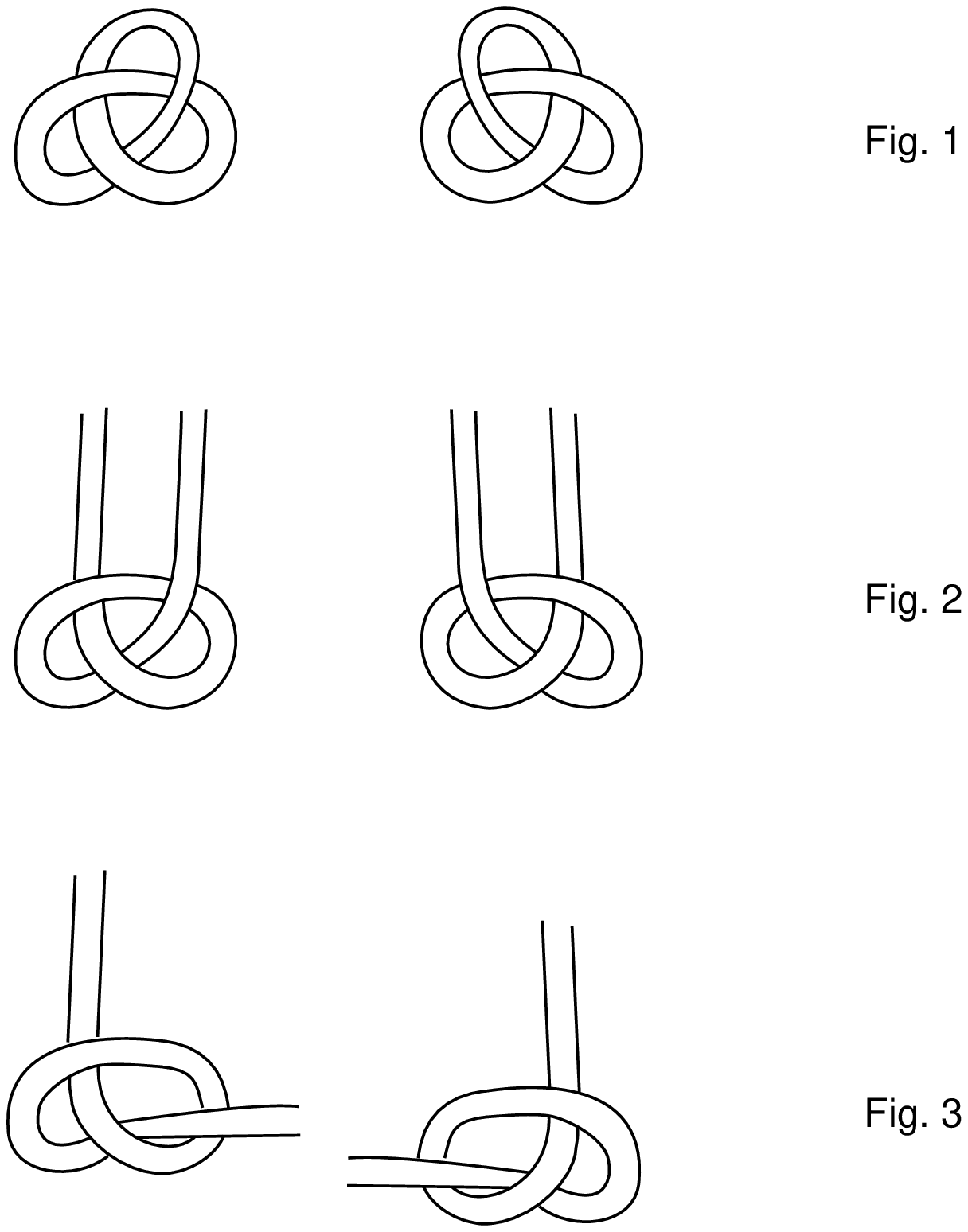,width=100mm,height=140mm}
}}
\end{picture}
\end{figure}
%

\newpage

\centerline{\;}

\begin{figure}
\centering
\setlength{\unitlength}{0.03mm}
\begin{picture}(2400,3400)
\put(-500,-1500){\makebox(0,0)[lb]{
\epsfig{file=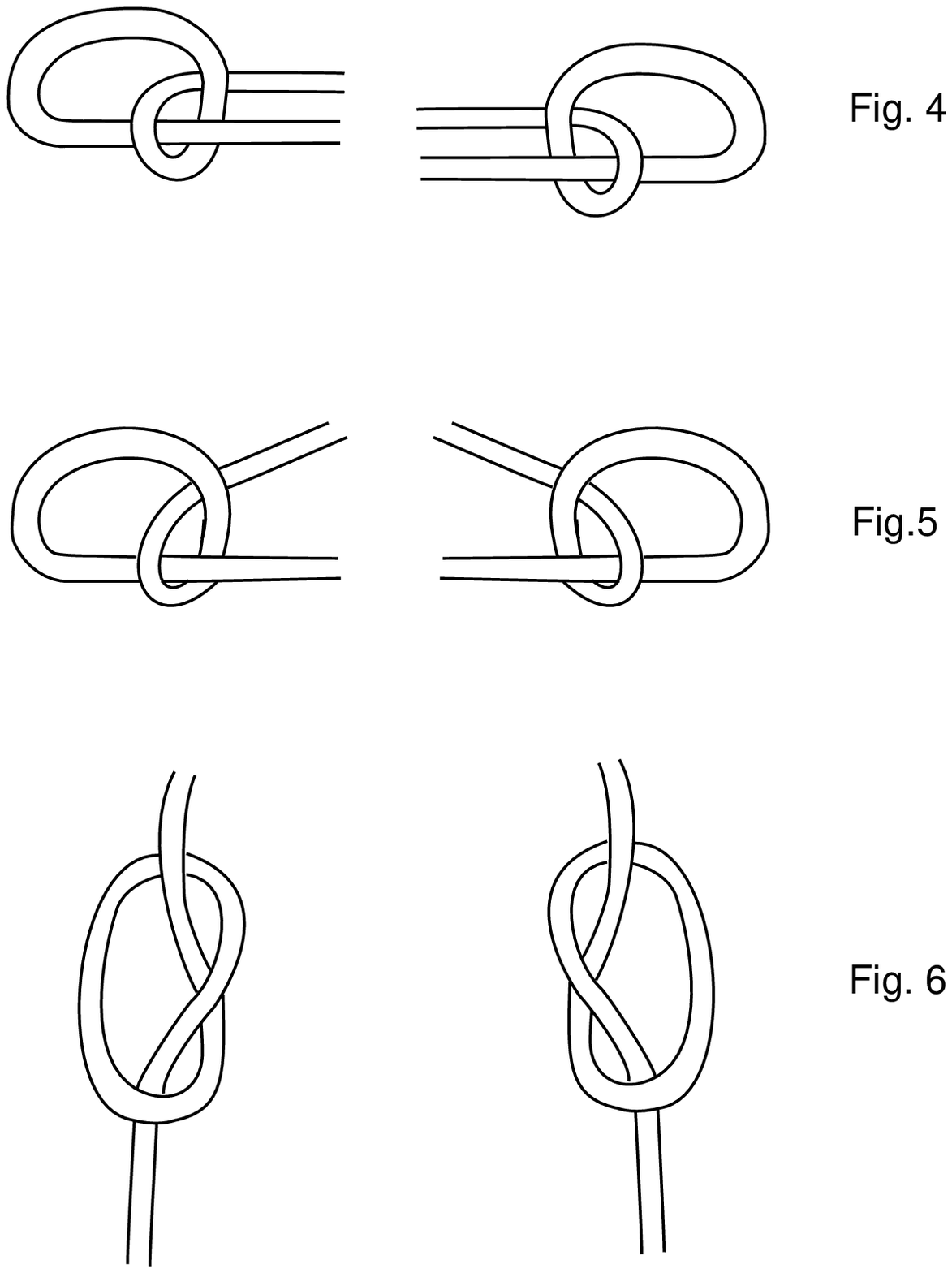,width=100mm,height=140mm}
}}
\end{picture}
\end{figure}


\end{document}